\documentclass[]{IEEEtran}
\usepackage{cite}
\usepackage{amsmath}
\usepackage{url}
\usepackage{flushend}
\usepackage{colortbl}
\usepackage{soul}
\usepackage{multirow}
\usepackage{pifont}
\usepackage{color}
\usepackage{alltt}
\usepackage[hidelinks]{hyperref}
\usepackage{enumerate}
\usepackage{siunitx}

\usepackage{pbox}
\usepackage{layouts} 
\usepackage[acronym]{glossaries}
\usepackage[capitalize]{cleveref} 
\usepackage[space]{grffile} 
\usepackage{subcaption}
\usepackage{bm} 
\usepackage{graphicx} 
\usepackage{epstopdf}
\usepackage{import} 

\usepackage{todonotes}
\usepackage{siunitx}
\usepackage{setspace}
\usepackage{enotez}

\mathcode`*=\string"8000
\begingroup
\catcode`*=\active
\xdef*{\noexpand\textup{\string*}}
\endgroup

\newacronym{AC}{AC}{alternating current}
\newacronym{DC}{DC}{direct current}
\newacronym{EV}{EV}{electric vehicle}
\newacronym{DSP}{DSP}{digital signal processor}
\newacronym{PWM}{PWM}{pulse width modulation}
\newacronym{THD}{THD}{total harmonic distortion}
\newacronym{SFDR}{SFDR}{spurious free dynamic range}
\newacronym{EMC}{EMC}{electromagnetic compatibility}
\newacronym{NPC}{NPC}{neutal point clamped}
\newacronym{HVDC}{HVDC}{high voltage direct current}
\newacronym{FET}{FET}{field-effect transistor}
\newacronym{IGBT}{IGBT}{insulated-gate bipolar transistor}
\newacronym{SM}{SM}{sub module}
\newacronym{DCDC}{DC-DC}{dc-to-dc}
\newacronym{SFT}{SFT}{sliding Fourier transform}
\newacronym{RMS}{RMS}{root mean squared}
\newacronym{BESS}{BESS}{battery energy storage system}
\newacronym{CHB}{CHB}{cascaded H-bridge}
\newacronym{MTPA}{MTPA}{maximum torque per ampere}
\newacronym{NEDC}{NEDC}{New European Drive Cycle}
\newacronym{WLTP}{WLTP}{Worldwide Harmonized Light-Duty Vehicles Test Procedure}
\newacronym{STATCOM}{STATCOM}{static synchronous compensator}
\newacronym{MMSPC}{MMSPC}{modular multilevel series parallel converter}
\newacronym{PCB}{PCB}{printed circuit board}
\newacronym{PMSM}{PMSM}{permanent magnet synchronous machine}
\newacronym{MMC}{MMC}{modular multilevel converter}
\newacronym{FPGA}{FPGA}{field-programmable gate array}
\newacronym{EMI}{EMI}{electromagnetic interference}
\newacronym{DFT}{DFT}{discrete Fourier transform}
\newacronym{FFT}{FFT}{fast Fourier transform}
\newacronym{GaN}{GaN}{gallium nitride}
\newacronym{SiC}{SiC}{silicon carbide}
\newacronym{SDM}{SDM}{Sigma-delta-modulation}
\newacronym{SINAD}{SINAD}{signal-to-noise and distortion ratio}
\newacronym{SoC}{SoC}{state of charge}
\newacronym{STFT}{STFT}{short-time Fourier transform}

\newcommand*\kc{\ensuremath{k}}

\newcommand*\Ss{\ensuremath{s}}
\newcommand*\Sw[1]{\ensuremath{S\left({#1}\right)}}
\newcommand*\N{\ensuremath{N}}
\newcommand*\Fw[1]{\ensuremath{F(\omega)_{#1}}}
\newcommand*\dft[1]{\ensuremath{\vec{D_\N\left({#1}\right)}}}
\newcommand*\eps[1]{\ensuremath{E_{#1}}}
\newcommand*\xn{\ensuremath{\vec{X_N}}}
\newcommand*\G{\ensuremath{\vec{G(\omega)}}}

\newcommand*\Vo{\ensuremath{V_\mathrm{o}}}

\newcommand*\Vi{\ensuremath{V_\mathrm{i}}}

\newcommand*\M{\ensuremath{M}}

\newcommand*\fc{\ensuremath{f_\mathrm{c}}}

\newcommand*\Km{\ensuremath{K_{\mathrm{max}}}}

\newcommand*\Jw[1]{\ensuremath{J_{#1}^1}}
\newcommand*\Js[1]{\ensuremath{J_{#1}^2}}
\newcommand*\Jr[1]{\ensuremath{J_{#1}^3}}

\newcommand*\diff{\mathop{}\!\mathrm{d}}

\renewcommand{\vec}[1]{\bm{#1}} 

\begin{document}
\title{Synchronous Multistep Predictive Spectral Control of the Switching Distortion in DC--DC Converters}

\author{
	\vspace{-0.1cm}
	{
	Christian Korte, \emph{Student Member}, \emph{IEEE}, Till Luetje, Stefan M. Goetz, \emph{Member}, \emph{IEEE}
	}
\vspace{-1cm}
}

\maketitle

\begin{abstract}
In automotive power electronics, distortion and electromagnetic interference (EMI) generated by the switching action of power semiconductors can be a significant challenge for the design of a compact, lightweight vehicle. As semiconductor switching frequencies increase, e.g., through the introduction of new materials, such as gallium nitride  and silicon carbide, this problem becomes more severe. We present a control scheme for an automotive dc-to-dc converter that reduces the EMI generated by shaping switching distortion predictively at the run time. The multistep model-predictive control scheme chooses the subsequent switching state that optimizes the output spectrum according to predefined criteria. To achieve real-time operation, it evaluates the possible switching state candidates for the next modulation step without explicitly solving a single Fourier transform.  {In addition, the switching rate and voltage ripple are controlled in a single unified control law. We present and experimentally validate that the control scheme can indeed run at real time already with currently available mid-range hardware. The results demonstrate that the largest spectral peak of the switching distortion can be decreased by 48~dB compared to conventional pulse-width modulation.}
Furthermore, spectral gaps can be implemented in the output distortion and altered in real-time, allowing certain frequency bands---e.g., bands used by other sensitive electronics such as sensors, communication busses, or tuners---to be kept free of EMI from the converter's switching actions. 

\end{abstract}


{}

\definecolor{limegreen}{rgb}{0.2, 0.8, 0.2}
\definecolor{forestgreen}{rgb}{0.13, 0.55, 0.13}
\definecolor{greenhtml}{rgb}{0.0, 0.5, 0.0}

\section{Introduction}
With the advancing electrification of modern vehicles---conventional, hybrid, and battery-powered---additional electrical and electronic systems emerge, and many previously purely mechanic units turn into electrical actuators, supplied through various voltage levels.
For many traditional low-power units, such as lights, control units, and entertainment systems, the conventional \SI{12}{\volt} supply is sufficient and established. Electric drives and high-power auxiliaries, such as compressors for the air conditioning and active chassis systems, however, would overwhelm the \SI{12}{\volt} auxiliary supply\cite{Evzelman2016}. Higher voltages, such as \SI{48}{\volt}, which is still within the safe extra-low voltage range with fewer risks for operators and technicians, and up to \SI{1,000}{\volt} are currently establishing to enable drive-train power levels and faster charging \cite{Kim2017,7873375,9343759}. 

\begin{figure}[t]
	\centering
	\includegraphics{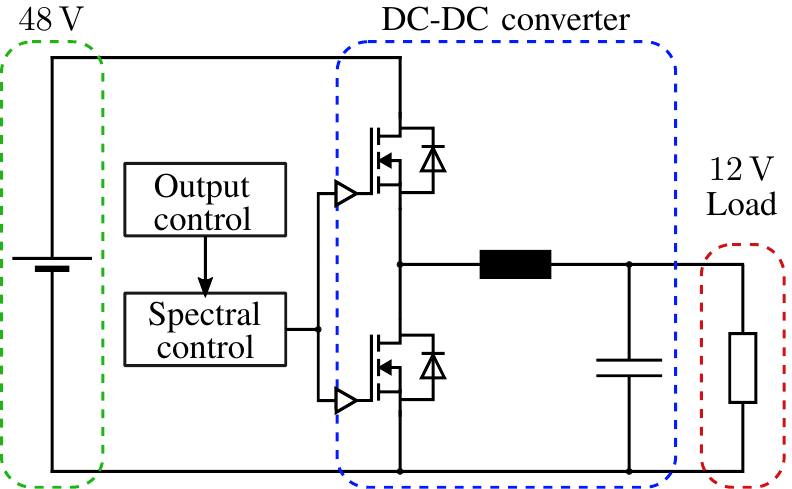}
	\caption{Synchronous buck converter with the spectral control scheme, which is composed of an output controller and a multistep predictive spectral controller.
	}
	\label{fig:DCDC}
\end{figure}

\let\colort\color
\begin{figure}
	\centering
	\includegraphics{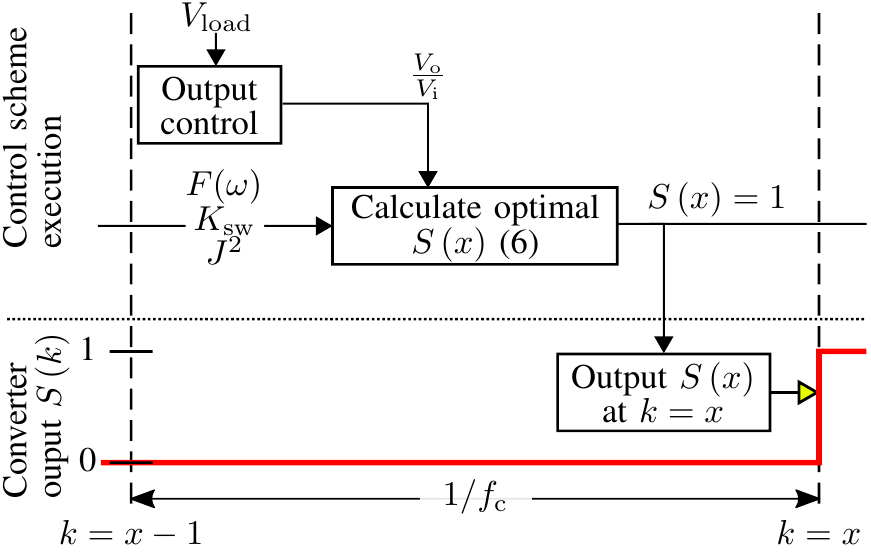}
	\caption{Flow diagram of the multistep predictive spectral control scheme showing the inputs and outputs for the two components of the controller. The voltage control controls the output voltage, while the calculation of (\ref{eq:costfun}) calculates the output state of the converter for the next control step. The period of the control scheme is $1/\fc$.
	}
	\label{fig:sw_zoom}
\end{figure}

Power converters are responsible for the power exchange between the various voltage levels and use switching modulation, typically pulse-width modulation (PWM) for a continuous control of the output.
Automotive \gls{DCDC} converters commonly switch at rates greater than \SI{50}{\kilo\hertz} to reduce the size of passive components \cite{Han2016}. 

As a result of the high-frequency switching action, the converters generate significant conducted and radiated \gls{EMI}. 
Aside from a improvement of silicon devices, the introduction of wide-bandgap semiconductors further aggravates the issue as they allow hard switching beyond 100~kHz, though typically not tolerated by car manufacturers in on-board applications due to the associated \gls{EMI} \cite{Liu2016,Han2017}. \gls{EMI} from power electronics can obstruct correct operation of other units in the vehicle and in extreme cases cause other systems such as field-bus systems or radio communications to fail\cite{Qu2018,Han2016}. In consequence, whereas wide-band-gap devices increase the technically achievable switching frequencies and converter power densities\cite{Zhang2015}, those developments do not reach the automotive world, which often even limit the allowable switching frequency due to EMI concerns.

Whereas conducted and radiated \gls{EMI} can occur at any frequency, the fixed constant switching frequency of \gls{PWM} concentrates the distortion power in narrow bands at the switching frequency and its harmonics\cite{Mainali2008}. These concentrated emissions conflict with typical \gls{EMI} standards used in automobiles such as CISPR 25\cite{cispr25}, which limit the spectral density.

\begin{figure*}
	\begin{center}
	\includegraphics{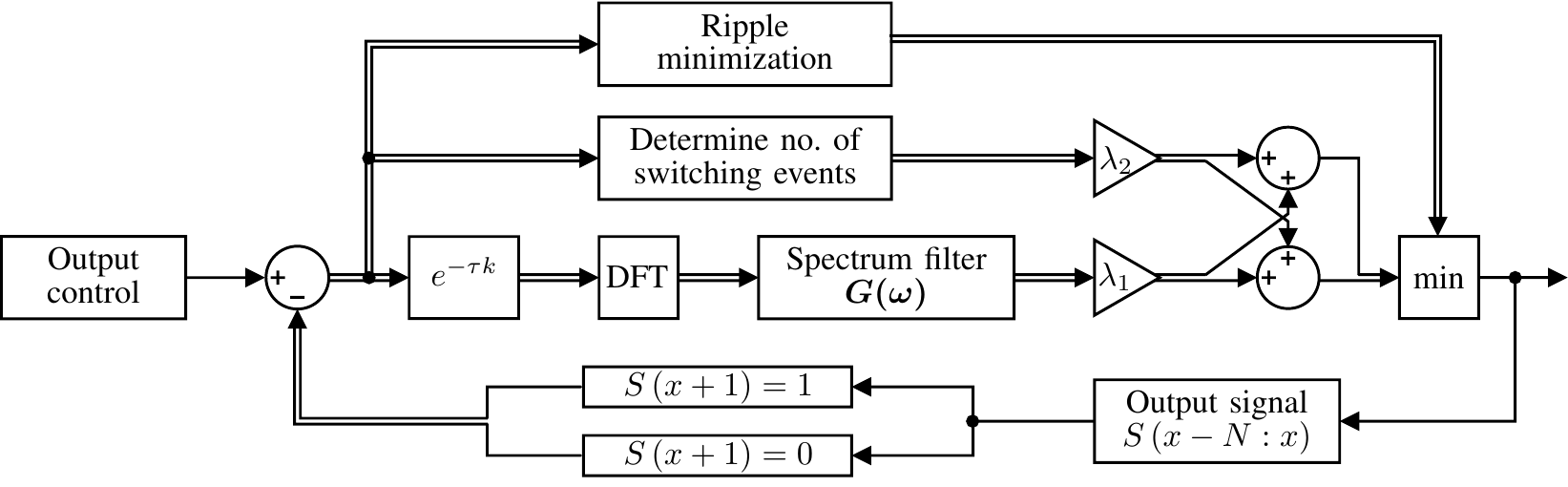}
	\end{center}
	\caption{Schematic of the calculations of the spectral control for the case $\M =1$. Two output options for $\Sw{x+1}$ are evaluated against three different criteria. The optimization function chooses the output with the minimum cost.}
	\label{fig:controlscheme}
\end{figure*}

Common remedies for emissions are spacious passive filters (capacitors and inductors) for conducted and radiated emissions and metallic shields to block radiated energy\cite{Ye2004}. All of these remedies, however, come at an increased hardware expense. Furthermore, filters often affect safety aspects, such as touch currents, and tend to interact with control \cite{6841996,6088240}.

{Multilevel circuit topologies can reduce EMI through a finer granularity of the switched voltage levels, and  for power inversion and, though rarer, also DC--DC conversion \cite{731062,1355682,8601394,7457710,9415177}. However, due to a higher complexity, typically more semiconductor devices, and larger size, multilevel converters to date practically do not play a role in automotive applications. Dynamically reconfigurable batteries, which exist for variable DC supply, with direct multiphase AC output, and also multiple outputs share large similarities with multilevel converters and may in the future but offer limited maturity yet \cite{9589040,7442763,7801087,8289145,arxiv.2202.11757,7468193}.}

Several methods have been proposed and investigated to reduce the \gls{EMI} at the source and consequently reduce the volume, weight, and cost necessary for additional hardware solutions by changing the modulation technique, typically to so-called spread-spectrum modulation\cite{Dousoky2010,Pareschi2015}. Spread-spectrum methods differ from regular \gls{PWM} in that the switching frequency of the semiconductors is changed to distribute the energy of the switching distortion to a greater frequency band, decreasing the spectral peaks. The switching frequency is either modulated in fixed patterns\cite{Tse2002,Yazdani2011}, or chaotically\cite{Mihalic2006,Jin2013}.

These methods suffer from the disadvantage that the time between two modulation intervals and therefore the timing of the control loop constantly change during operation as the switching frequency changes\cite{Pareschi2015}. The changing timing complicates the control and renders the implementation difficult in systems which generally rely on fixed cycle times. Furthermore, as the control-loop typically requires a minimum amount of time to compute, the switching frequency can only be spread to lower frequencies. This can reduce the performance of the converter in terms of output current and voltage ripple. 

\gls{SDM} has been proposed to reduce the \gls{EMI} of \gls{DCDC} converters\cite{Yan2011,Hwang2016}, although \gls{SDM} can result in load-dependent switching frequencies and suffers from stability issues when it is of higher order. The distortion spectrum can also not be shaped according to a predefined target shape to avoid specific frequencies at which other components may be highly susceptible to \gls{EMI}.

In this contribution, we analyze in detail and extend previously proposed predictive spectral control strategies that control the \gls{EMI} of a \gls{DCDC} converter according to a predefined spectrum by altering the spectrum of the transistor switching signals \cite{Goetz2016,Korte2017,8099351,8289243}. {This contribution enables and studies prediction horizons of the spectral control scheme up to eight steps into the future. A comprehensive guideline for implementing the spectral control scheme is presented, including the details on how the control equations can be implemented on a \gls{FPGA}. We furthermore present the ability of the control scheme to adapt to changes in the demanded spectrum shape and limit the converter output ripple in the required spectrum.}
We will demonstrate that this multistep predictive method already allows real-time operation. With growing speed and number of processing units or logic blocks of signal processors and FPGAs, the achievable switching speed will continually increase.

This method can systematically reduce the emission and even design as well as synthesize the distortion spectrum of conventional silicon \gls{DCDC} converters as well as wide-bandgap circuits.
The spectral control method can determine in which frequency band \gls{EMI} is tolerated and in which bands it should be prevented. Furthermore, the control strategy incorporates additional target functions that ensure the output voltage ripple and the average switching frequency of the converter can be directly controlled.

The proposed method controls the output spectrum of the \gls{DCDC} converter by calculating the spectrum of the past switching signals in each control cycle and choosing the switching state for the next output cycle that results in the optimum spectrum according to a desired spectral shape.

For the purpose of a general analysis, we use a synchronous buck converter transferring power between a \SI{48}{\volt} energy storage device and a \SI{12}{\volt} auxiliary supply to implement and evaluate the presented control scheme, as shown in \cref{fig:DCDC}.

\section{Control Approach}
\label{sec:cntrlapp}

In this section, we introduce the novel predictive spectral shaping method as a comprehensive approach to controlling the output distortion of a buck converter, allowing the user to shape the spectrum at run time.

\subsection{Model Predictive Spectral Shaping}

The presented model-predictive control strategy evaluates a cost function for all possible subsequent switching state candidates of the converter to directly determine the transistors' switching signals of the DC-DC converter. {As such it is a direct control strategy that does not require a subsequent modulator and is not based on techniques such as frequency modulated \gls{PWM}, which alter the modulation in order to affect the output spectrum.}
{The duty cycle of the converter as well as the average switching frequency are therefore not inputs to the control scheme but are results from the switching signals generated by the model-predictive optimization function, and difficult to predict accurately in advance.}

While the presented scheme can be extended to include any number of criteria, this paper focuses on three: shaping of the output distortion spectrum, control of the average switching rate, and reduction of the output voltage ripple. These are individually discussed and presented in the following subsections.

\subsubsection{Spectral Control}

The main focus of the novel control method is the spectral control, which allows the user to affect the output spectrum of the converter based on requirements such as resonances or particularly \gls{EMI}.

{The output voltage power spectrum of the converter is estimated by calculating the spectrum of the \gls{FET} switching-signal $\vec{\Sw{k}}$ (shown in \Cref{fig:timesim}) for the next \M\ switching steps for all possible switching states of the converter, where \M\ is the prediction horizon, or number of controller steps into the future for which the control scheme calculates the cost function.}
For a synchronous buck converter, there are two possible switching states $\Sw{\kc}\in\left\{0,1\right\}$ for each control step \kc. {Therefore the number of options that are evaluated at each control cycle \kc\ is $2^\M$.}
The output state of the power switches is optimized in each control cycle.

{Increasing the prediction horizon allows the control scheme to more closely follow the desired output spectrum with fewer switching actions, as the evaluated switching possibilities increase exponentially with the prediction horizon. Obviously, this also results in an exponentially increasing computational power demand.}

A spectral analysis, such as \gls{DFT}, of the switching signal for all possible candidate states 
\begin{equation}
\label{eq:ffts}
\vec{\Fw{\Ss}}=\mathcal{F}\left(\vec{\Sw{k}\right})
\end{equation}
can provide the corresponding spectra $\vec{\Fw{\Ss}}$.
The time period of the \gls{DFT} calculation is \M\ steps in the future ($\kc+\M$) to a certain number of steps in the past ($\kc-\N+\M$), where \N\ is the number of \gls{DFT} samples. {Each control step corresponds to one sample point of $\vec{\Sw{\kc}}$, therefore the maximum frequency the control scheme can influence is half the control frequency \fc, as shown in \cref{fig:sw_zoom}. The transistors are assumed to switch instantaneously between successive control steps so that the dead time is not immediately incorporated but could use known compensation methods that act on the transition timings or the output control loop can compensate the amplitude, whereas the spectral effect will be left to the spectral control \cite{774205,564157}.}
\cref{sec:implement} provides further details on the implementation of the calculation.

The spectrum of the switching signal is evaluated against a reference spectrum $F^{\ast}(\omega)$ according to
\begin{equation}
\label{eq:ffterr}
\eps{\Ss}=\int{\vec{\Fw{\Ss}}-\vec{F^{\ast}(\omega)}\diff \omega}.
\end{equation}
\eps{\Ss} evaluates the quality of the output spectrum for the possible switching states.

If the DC component $F(0)$ of the spectrum is ignored, the reference spectrum can be set to zero ($\vec{F^{\ast}(\omega)}=0$). This reflects that for a DC-DC converter the only desirable energy is at frequency zero. The implementation of the desired value of the DC component is given in \cref{sec:dccomp}.

\Cref{eq:ffterr} calculates the quality of the output spectrum, and is multiplied by a spectral filter $\G$, which allows specific parts of the distortion spectrum to be weighted more than others. The spectral cost $\Jw{\Ss}$ of the output options is therefore given by
\begin{equation}
\label{eq:fftcost}
\Jw{\Ss}=\left\lVert\vec{\G}\odot\left(\vec{\Fw{\Ss}}-\vec{F^{\ast}(\omega)}\right)\right\rVert^p
\end{equation}
where $\odot$ denotes the Hadamard product.

{For the selected case when $\vec{\G} = 0$, the control scheme choses the switching states of the controller that correspond most closely to the desired spectrum $F^{\ast}(\omega)$, which is a spectrum with no peaks and the switching energy spread across the entire frequency range.}

The p-norm allows the cost of the spectrum to be tuned, for instance, to minimize the largest peaks ($p \to \inf$), the overall power ($p  = 2$) or the amplitude ($p  = 1$) of the output spectrum distortion.

In addition to the cost of the output spectrum, the presented control scheme evaluates two further cost contributors: the cost of switching and the cost of the output ripple.

\subsubsection{Switching Rate Reduction}

The cost of switching \Js{\Ss} is determined simply by cumulating the number of switching events in the time-window

\begin{equation}
\label{eq:swcost}
\Js{\Ss} = \sum\limits_{\kc=1}^{\N-1} \Sw{\kc} \neq \Sw{\kc+1}.
\end{equation}

As the Fourier transform is always performed over the same number of samples -- and therefore temporal window length -- this value is proportional to the average switching frequency of the converter. Increasing the weight of the cost \Js{}\ therefore reduces the average switching frequency of the converter.

Reducing the number of switching instances {can allow the converter to reduce the switching losses without directly measuring them, as they are approximately a linear function of the switching rate in DC applications. Knowledge of the output voltage and current could be used to directly reduce the switching losses, for example by increasing the weight of the switching penalty when the current is high.}

The reduction of the switching rate will not lead to further energy in the switching distortion, however it may impede the ability of the spectral control scheme to adhere to the target spectrum.

{ The maximum instantaneous switching frequency of the spectral control system is half of the control frequency, as the transistor states only change at the end of a control step. This ensures that the sampling frequency is always at least twice as high as the switching frequency.}

\subsubsection{Ripple Minimization}

While the reduction of the switching rate can have beneficial properties, a longer time period between switching instances will increase the instantaneous voltage ripple. {An increase in voltage ripple leads to higher conduction losses, higher maximum transistor currents and can lead to disturbances of system loads.}
To counteract this, a voltage ripple reduction is added to the model-predictive framework to control the tolerable voltage ripple. The cost of the time since the last switching instance, \Jr{}, is given by
\begin{equation}
\label{eq:vrcost}
\Jr{\Ss} = \begin{cases}
0 & K_\mathrm{sw} > K_\mathrm{max}\\
-\infty & K_\mathrm{sw} = K_\mathrm{max},
\end{cases}
\end{equation}
where $K_{sw}$ is the number of consecutive past control cycles for which no switching action has taken place, and $K_\mathrm{max}$ is the maximum tolerated number of control cycles in which no switching may take place, and is chosen by the user. {Assuming a constant output voltage (negligible voltage ripple relative to the output voltage), the voltage ripple is proportional to the maximum time between two switching actions. With the knowledge of the \gls{DCDC} inductance and output capacitance, the output voltage ripple can thus be estimated and limited based on a simple inductor--capacitor model, without directly measuring the ripple.}

\subsection{Cost Function}

In order to apply the optimal output in the next switching instance, according to the three presented cost terms, a cost function weights the possible options per
\begin{equation}
\label{eq:costfun}
\Sw{k+1}=\textrm{arg}\text{min}(\lambda_1\Jw{\Ss} + \lambda_2\Js{\Ss} + \Jr{\Ss}),
\end{equation}
where $\lambda_\mathrm{i}$ is the weight of each associated cost. {\cref{fig:controlscheme} provides a visual overview of how the different cost terms for the cost function are calculated an weighted, based on the two possible outputs of the control scheme when $\M=1$.}
In order to tune the controller, the three weights can be altered to balance the control scheme towards a more optimized output spectrum, a lower switching rate or a lower voltage ripple. 
The weights may also be altered during run-time, for example to have a larger focus on switching rate and therefore efficiency during operation at high loads, and a more optimized spectrum when the losses are lower.

\subsection{Output Control}

While the presented model-predictive control generates the desired output spectrum of $\vec{\Sw{k}}$, it has no feedback loop to accurately and dynamically control the output voltage of the converter.

For this reason, an output voltage feedback and/or feed-forward controller can be used to regulate the DC-DC converter output.
The structure of the control scheme can allow any form of output control to be used in conjunction with the spectral control, and for details the authors refer to abundant literature on the subject\cite{Babaa1966,Park2014}.

\subsubsection{DC Component}
\label{sec:dccomp}
The cascaded output controller is used to alter the DC offset in the calculation of the \gls{DFT}, such that the evaluation of the spectrum only requires the analysis of the non-zero frequency components of the output signal. This allows the zero-frequency of the reference spectrum $F^{\ast}(0)$ to be set to zero, simplifying the implementation of the filter $\vec{\G}$.

In order for the zero-frequency component of a spectrum to be zero, the mean value of the signal must be zero. 
Therefore, the output switching signal $\vec{\Sw{\kc}}$ is shifted by the ratio of the reference voltage of the output control \Vo\ to the input voltage of the converter \Vi, so that $\tilde{\Sw{\kc}}\in\left\{-\frac{\Vo}{\Vi},1-\frac{\Vo}{\Vi}\right\}$ instead of $\Sw{\kc}\in\left\{0,1\right\}$ for the calculation of $\vec{\Fw{\Ss}}$.

This shift ensures that the output voltage follows the set-point given by the outer control loop, without evaluating it in the spectral cost.

\section{Controller Design and Implementation} 
\label{sec:implement}

This section provides a guide for the design of the multistep predictive spectral controller and details on how to enable the implementation on an \gls{FPGA}.

\subsubsection{Controller Design}

{
Since two cascaded controllers---the output control and the spectral control---form the presented control system, the interaction between both controllers must be considered during all steps of the controller design. As the focus of this paper is the spectral control, the design of the output control is not treated here in detail.

Due to the cascaded nature of the control scheme, the outer voltage controller must have a significantly lower than the inner spectral control. For the design of the voltage controller, the spectral control is modeled as a 1\textsuperscript{st}-order delay element with a time constant of $5/(2\pi\fc)$, which approximately corresponds to the average switching frequency generated by the spectral control for the case $\lambda_2=0$.
While the target switching frequency determines the controller frequency, the number of samples used for the time-window of the Fourier transform governs the frequency resolution of the output spectrum. A sample number of $\N\geq2000$ has been found to allow good spectral shaping, while using too few samples can cause peaks in the output spectrum.

The output voltage controller can be designed based on this simplified model of the spectral control. Knowledge of the output voltage controller can then be used to tune the design of the spectral control to ensure that each controller's actions do not interfere with the control target of the respective other controller. This interaction is suppressed by the filter \G. The value of \G\ must be large enough in the frequency range in which the controller is active (\SI{0}{\hertz} to $\fc/10$), as this ensures that the subsequent spectral control adheres to the output of the controller. Furthermore, the stability of the spectral control is improved by making the filter magnitude inversely proportional to the frequency (see \cref{fig:specgap}).

Given that the lower frequencies of the filter \G\ allow the output control to regulate the voltage, the higher frequencies ($\fc/10$ to $\fc/2$) can be used to shape the spectral distortion generated by the switching of the semiconductors, where the upper limit $\fc/2$ arises from the Nyquist–Shannon sampling theorem\cite{Shannon}.

In this frequency region, the output distortion can be shaped to fulfill the \gls{EMI} targets of the application. Targets can, for example, be that the distortion energy is spread over the entire spectrum; that certain ranges of the spectrum are free from distortion; or that the minimum frequency of the distortion or that peaks in the distortion are avoided.

Sections of \G\ that are proportional to the frequency suppress tones in the controller which otherwise caused spectral peaks, where distortion energy is concentrated around specific frequencies. The weight rising linearly with the frequency causes the spectral control to utilize more of the low-frequency spectrum and hence distributes the distortion energy. 

In addition to spreading the distortion energy, the minimum frequency of the distortion can be an important parameter of the \gls{EMI} performance. The minimum distortion frequency is predominantly controlled by the frequency above the PI controller's bandwidth. At this frequency $\fc/10$, there is a sharp decrease of \G, which can be seen in \cref{fig:specgap}.

In addition to spectral spreading and peak reduction, \gls{EMI} requirements such as conducted emissions targets can require some frequencies to contain less distortion than other nearby frequencies. Such spectral valleys can be achieved using spectral control by adding notches to the filter \G\ where the \gls{EMI} should be avoided. It should be noted that the presented spectral control method directly optimizes the spectrum of the switching signal of the transistors, not of the output voltage of the converter system. In order to determine the amount of conducted and radiated \gls{EMI} generated by the converter system, the output inductor and capacitor need to be taken into account as well as filters, which are generally used in automotive power electronics. Furthermore, current path routing, shielding and the behavior of the power source and load can have a significant impact on the interference generated.

While the entire \gls{EMC} design of an automotive converter is out-of-scope for this contribution, the presented spectral control scheme provides a tool for engineers to optimize \gls{DCDC}-converter \gls{EMI} at the source, the switching of the power semiconductors.

Once the output spectrum generated by the control scheme is satisfactory, the further cost terms \Js{} and \Jr{} can be included to adjust the switching frequency and output ripple to the requirements of the application. At first, the desired average switching frequency is obtained by increasing $\lambda_1$. Secondly, the output voltage ripple can be limited by choosing $K_\mathrm{max}$.

}
Substantial computational resources are required to compute predictive spectral control. The artifices that allow computation at run time are discussed in the following.

\subsubsection{Discrete Fourier Transform Calculation}

The most computationally intensive aspect of the spectral control is the calculation of the Fourier transform of the output switching signal. In principle, the frequency representation has to be calculated over the number of samples in the time window at least twice at each modulation step for a prediction horizon of $\M=1$.

We derive the spectrum at the present time and into the future from the previous one without explicitly solving a single Fourier transform. This stunt exploits two features of the frequency domain:
(1) time shifts, e.g., due to the shifting \textit{now} with the progressing time, corresponds to phase shifts in the spectrum;
(2) the approximately rectangular voltage and triangular current response to switching of the state candidates in the future can be described and evaluated analytically \cite{Jacobsen2004}. The computational burden is substantially reduced even compared to \gls{FFT} methods, requires moderate resources and especially no iteration on an \gls{FPGA}, and increases only linearly with the window size ($o(N)$)---compared to order $N \log{N}$ for an \gls{FFT} and $N^2$ for a full discrete Fourier transform. Still, ignoring considering numerical errors, such progressive semi-analytical evaluation of the spectrum is mathematically equivalent to a full numerical discrete Fourier transform.


Accordingly, we calculate the spectrum including the current modulation step \dft{\kc}, using the next (potential) sample of the switching vector \Sw{\kc+1}, the oldest sample of the switching vector \Sw{\kc-\N} and a shift vector operator \xn, defined as
\begin{equation}
\label{eq:shftvec}
\xn=\exp{\frac{j2\pi \vec{n}}{\N}},
\end{equation}
where $\vec{n} = \left[0 ... \N-1\right]$. The spectrum of the current modulation step is then calculated according to
\begin{equation}
\label{eq:sft}
\dft{k}=\left(\dft{k-1}-\Sw{\kc-\N}+\Sw{\kc}\right)\odot\xn.
\end{equation}
Between consecutive control steps, only the vectors \dft{k}\ and \xn\ need to be stored.

\begin{figure}
	\centering
    \includegraphics{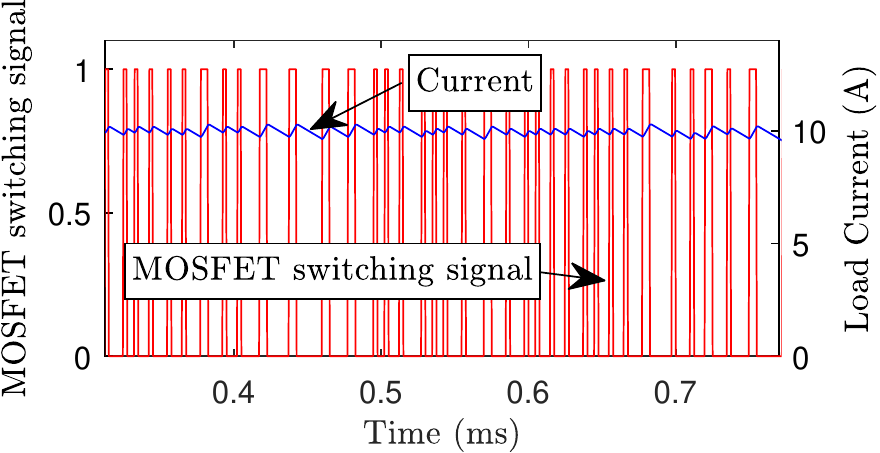}
    \caption{Output current and switching waveforms of the presented spectral control method. The control frequency \fc\ is \SI{400}{\kilo\hertz}, resulting in an average switching frequency of \SI{77}{\kilo\hertz}.}
	\label{fig:timesim}
\end{figure}

\begin{figure}
	\centering
    \includegraphics{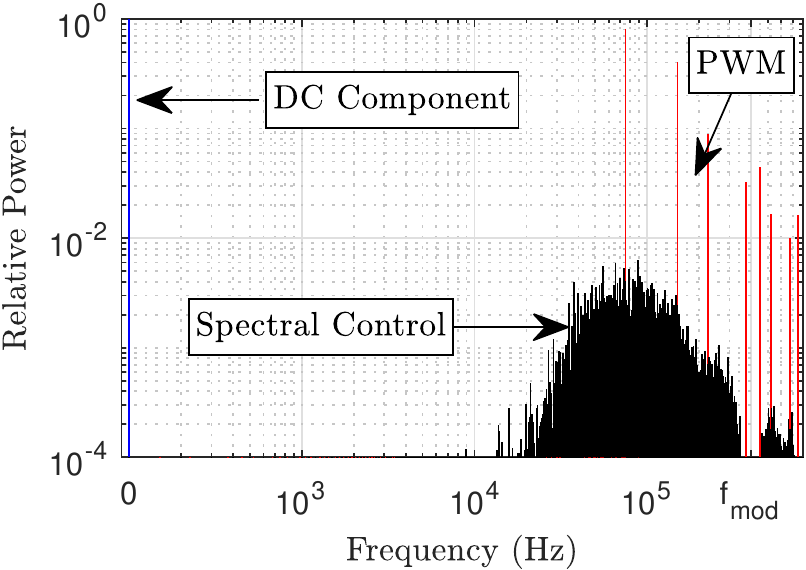}
	\caption{Output spectrum of the presented spectral control scheme compared to PWM at the same average switching frequency. $f_{\mathrm{mod}}$ indicates the update rate, while the power of the \SI{0}{\hertz} component is added for reference. In contrast to \gls{PWM}, the spectral control can spread the switching distortion across a wide frequency range an avoid peaks in the spectrum.}
	\label{fig:spectime}
\end{figure}

\section{Model}

The proposed control method was calibrated in MATLAB/Simulink on a buck converter with an inductance of \SI{42}{\micro\henry} and an output capacitance of \SI{5000}{\micro\farad}. The control scheme runs at a constant frequency of \SI{400}{\kilo\hertz}, with an input voltage of \SI{48}{\volt} and an output voltage of \SI{12}{\volt}.

In \cref{fig:timesim} the output current of the converter and the switching signal of the high-side \gls{FET} is shown. The relative output power spectrum of the switching signal, normalized to the power of the \SI{0}{\hertz} component, is shown in \cref{fig:spectime}. {The switching actions are not spaced regularly but are spread out by the control to follow its target shape. This allows any peaks in the output spectrum to be avoided.}

For this simulation, the prediction horizon of the control scheme looks two steps into the future, and the p-norm is infinite. As a result, spectral peaks in the output distortion are minimized. Furthermore, the weight for the average switching rate $\lambda_2$ is set to zero while the voltage ripple control is not used.

\begin{figure}
	\centering
    \includegraphics{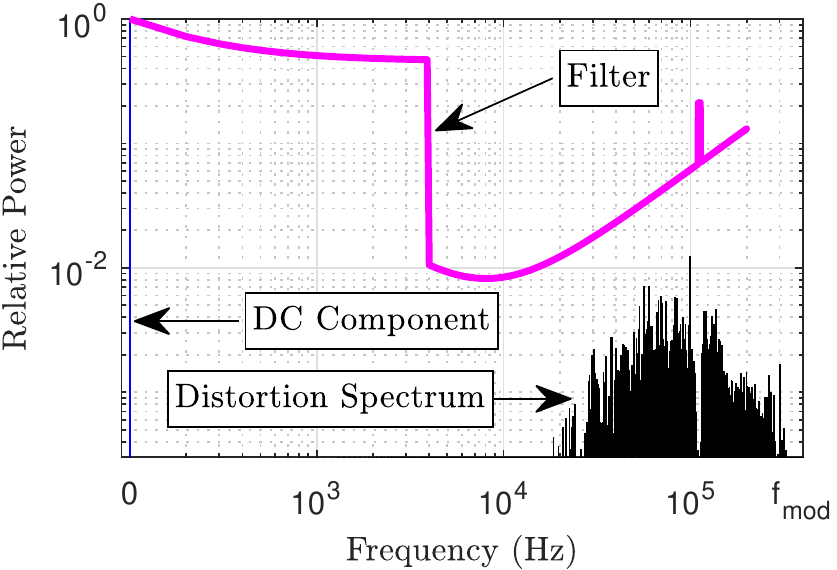}
	\caption{Output spectrum when using spectral control to implement a noise gap between \SI{99}{\kilo\hertz} and \SI{101}{\kilo\hertz}. The filter \G\ used to generate the spectrum is also shown. Note the high value of the filter at frequencies below \SI{99}{\kilo\hertz}, which ensures the output of the voltage controller is adhered to.}
	\label{fig:specgap}
\end{figure}

The spectrum in \cref{fig:spectime} shows that the spectral control evenly distributes the inevitable distortion generated by the switching action of the converter. In comparison, conventional \gls{PWM} at a similar average switching frequency (\SI{75}{\kilo\hertz} for \gls{PWM} compared to \SI{77}{\kilo\hertz} for the spectral control) has the distortion power concentrated around the switching frequency and its harmonics. {The spectral control achieves a \gls{SFDR} of \SI{22.0}{\decibel}, while the conventional \gls{PWM} signal has an \gls{SFDR} of \SI{0.91}{\decibel}.}
Since \gls{EMI} is created by the energy at a specific frequency, the interference is greatly reduced using spectral control.

{While the output spectrum is improved compared to conventional \gls{PWM} modulation, the output voltage ripple is significantly worsened. When switching from the spectral control to \gls{PWM}, the peak-to-peak ripple is increased by a factor of 2.7, the variance of the output voltage by 2.6, practically as the cost of the spectral flexibility.
It should be noted, however, that a frequency modulation technique achieving the same spectral spreading will also increase the voltage ripple compared to the best case, which is single frequency \gls{PWM}.}

In \cref{fig:specgap}, the filter \G\ is used to create a spectral gap in the switching distortion. While the remaining conditions are kept the same, the spectral gap is generated to suppress the switching distortion in the frequency band between \SI{99}{\kilo\hertz} and \SI{101}{\kilo\hertz}. This can be used to reduce interference at a specific frequency. As a result of the spectral gap, the remainder of the distortion spectrum increases, absoring the power previously in the gap. This icrease can be seen from the larger spectral spikes when compared to \cref{fig:timesim}, slightly reducing the \gls{SFDR} by \SI{3}{\decibel}.

\begin{figure}
	\centering
	\begin{subfigure}[c]{\columnwidth}
		\centering
		\includegraphics{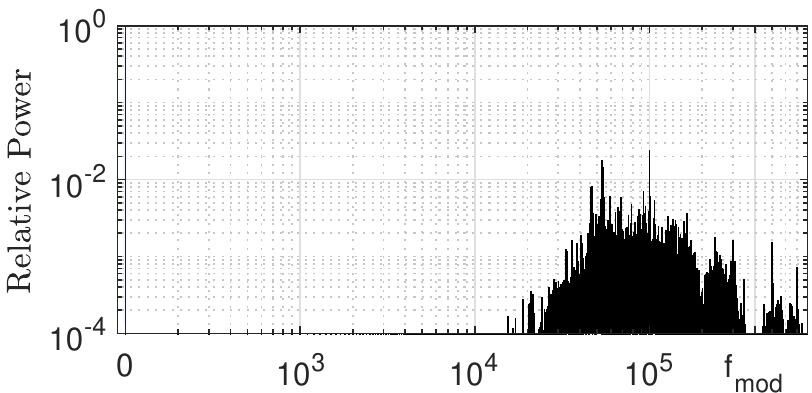}
		\subcaption{$\M = 1$}
		\label{fig:mstep1}	
	\end{subfigure}
	\begin{subfigure}[c]{\columnwidth}
		\centering
		\includegraphics{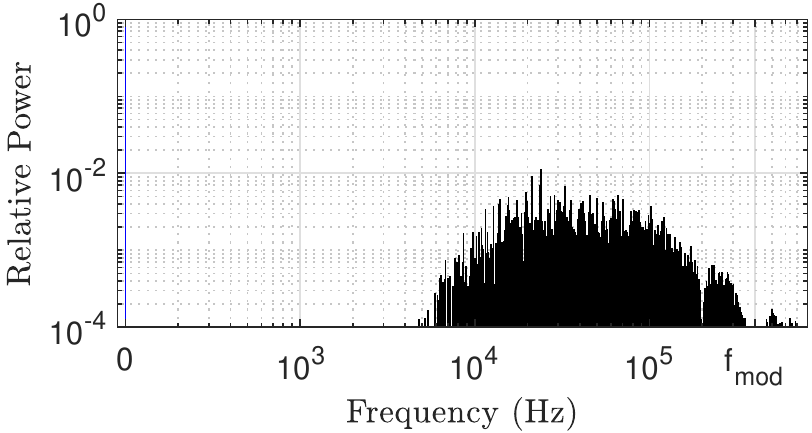}
		\subcaption{$\M = 8$}
		\label{fig:mstep10}	
	\end{subfigure}
	\caption{Output spectrum of the control scheme for prediction horizons $\M = 1$ and $\M = 8$. Increasing the prediction horizon allows the controller to more closely follow the reference spectrum and spreads the distortion more evenly.}
	\label{fig:mstep}
\end{figure}

\begin{figure}
	\centering
	\includegraphics{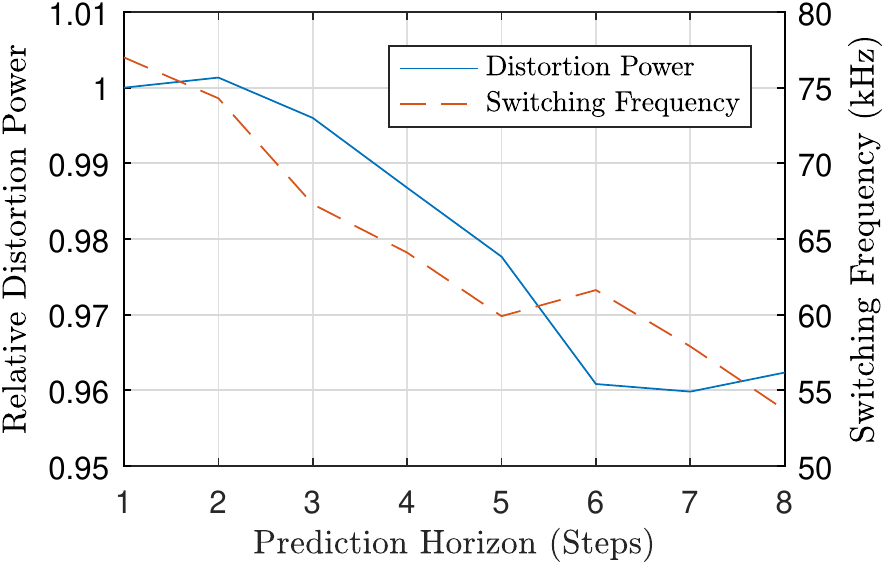}
	\caption{Distortion Power and average switching frequency plotted over the prediction horizon. As the prediction horizon increases, both the distortion and average switching frequency decrease as the controller can optimize several steps in advance.}
	\label{fig:statsmstp}
\end{figure}

The prediction horizon of the model-predictive control scheme can be increased to further optimize the distortion spectrum generated by the spectral control. This however comes at the cost of increasing computational power required.

Simulations were performed using the spectral control algorithm with prediction horizons of 1 to 8 steps. All other conditions are kept the same. The reference spectrum is a flat distortion distribution from \SI{10}{\kilo\hertz} to half the update frequency (\SI{200}{\kilo\hertz}), with a gap in the distortion around \SI{200}{\kilo\hertz}.

The increase of the prediction horizon allows the control scheme to more accurately shape the distortion spectrum according to the demand, as seen in \cref{fig:mstep}. An eight-fold increase of the prediction horizon allows the largest spurious spike in the distortion to be reduced by a factor of three, and the shape of the noise is flatter across the permitted noise frequencies. Furthermore, the spectral gap at \SI{200}{\kilo\hertz} is more pronounced.

\cref{fig:statsmstp} shows relevant performance parameters of the spectral control as the number of steps in the prediction horizon is increased. It can be seen that both the average switching rate and the power of the output distortion decrease as the number of prediction steps is increased. The average switching frequency can be reduced simply by increasing the prediction horizon as this allows the control scheme to modify the switching pattern several steps into the future.

\begin{figure}
	\centering
	\includegraphics[scale=0.85]{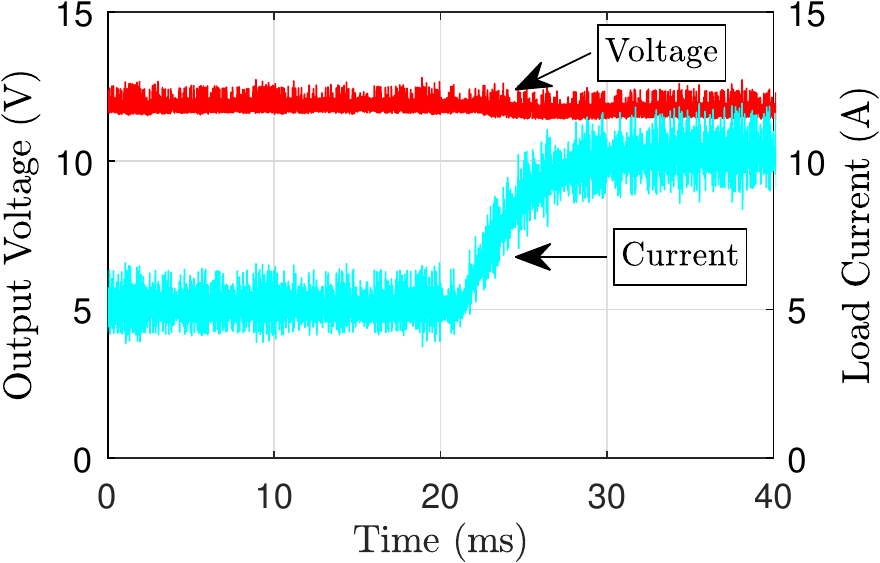}
	\caption{Output current and voltage of the experimental prototype during a load step from \SI{5}{\ampere} to \SI{10}{\ampere}. The controller is able to maintain a constant voltage while generating a distortion spectrum according to the filter \G\ (shown in  \cref{fig:timespecmeas}).}
	\label{fig:timemeas}
\end{figure}

\section{Experimental Validation}

\begin{figure}
	\centering
	\begin{subfigure}[c]{\columnwidth}
		\centering
		\includegraphics{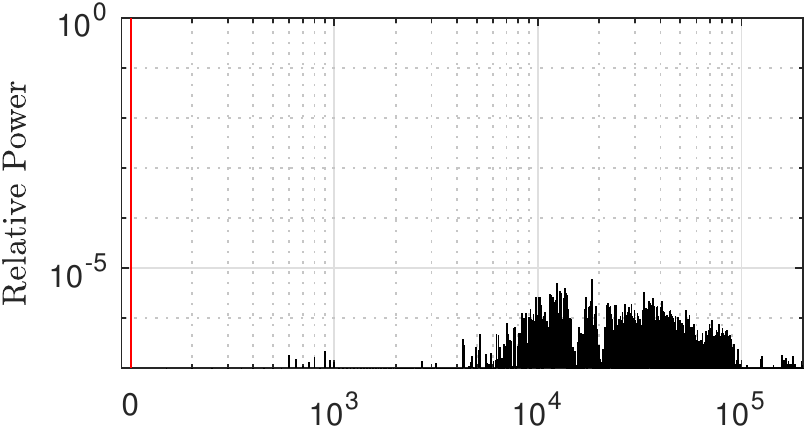}
		\subcaption{Before the load step}
		\label{fig:timespecmeasbef}	
	\end{subfigure}\\
	\begin{subfigure}[c]{\columnwidth}
		\centering
		\includegraphics{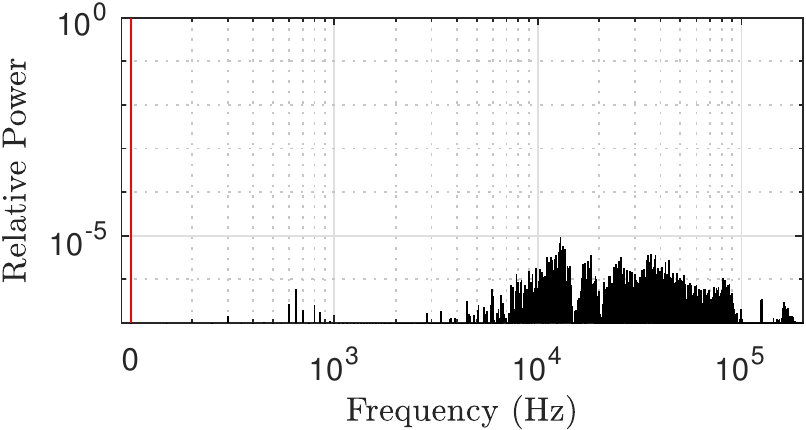}
		\subcaption{After the load step}
		\label{fig:timespecmeasaft}	
	\end{subfigure}
	\caption{Output distortion spectra of the converter prototype before and after the load step shown in \cref{fig:timemeas}. Two noise gaps are generated by the spectral control scheme, at \SI{15}{\kilo\hertz} and \SI{20}{\kilo\hertz}, respectively.}
	\label{fig:timespecmeas}
\end{figure}

\subsection{Setup}

We demonstrate the ability to successfully compute the spectral control method on a presently available IC at run time. An increase in computational power promises faster operation and an increase of the controllable spectral bandwidth.
The control scheme was verified experimentally on a synchronous buck converter, with an input voltage of \SI{48}{\volt} and an output voltage of \SI{12}{\volt}. The converter uses fast field-effect transistors (GS66508T, Gan Systems, Montreal, Canada). {It should be noted that for the output spectrum calculation the switching is assumed to occur instantly. Switching transitions are not optimized by the controller as they affect the \gls{EMI} at frequencies far above the controller's Nyquist frequency.}
The converter is operated without output filter to minimize modification of the distortion.

\cref{tab:convstats} summarizes the most important parameters of the developed converter. A power supply EA PS8160-170 and a load EA EL9080-170B (both from Elektro-Automatik, Viersen, Germany) serve as the source and sink for the converter current, respectively.

A medium-level Kintex-7 \gls{FPGA} (Xilinx, San Jose, CA) on a compact RIO 9039 (National Instruments, Austin, TX) calculates the control scheme in real time. The calculations are performed in 32-bit depth using a time window of 2047 samples at a control frequency of \SI{125}{\kilo\hertz}. Future \gls{FPGA}s or digital signal processors (DSP) will sucessively increase the achievable control frequency. An outer PI controller regulates the output voltage. Due to the computational effort, the control scheme uses a prediction horizon of one is used for subsequent measurements.

\begin{table}
	\centering
	\caption{Parameters of the buck converter used to verify the spectral control method	
	}
	\label{tab:convstats}
	\begin{tabular}{l|c}
		Parameter             	& Value                 \\
		\hline
		Control frequency  	& \SI{125}{\kilo\hertz} \\
		Max output current           	& \SI{30}{\ampere}      \\
		\gls{DCDC}-inductance 	& \SI{22}{\micro\henry} \\
		Input capacitance     	& \SI{5}{\milli\farad}  \\
		Input voltage range		& \SIrange{34}{60}{\volt}\\
		Output capacitance    	& \SI{15}{\micro\farad}	\\
		Output power			& \SI{300}{\watt}	
	\end{tabular}
\end{table}

\begin{figure}
	\centering
	\includegraphics{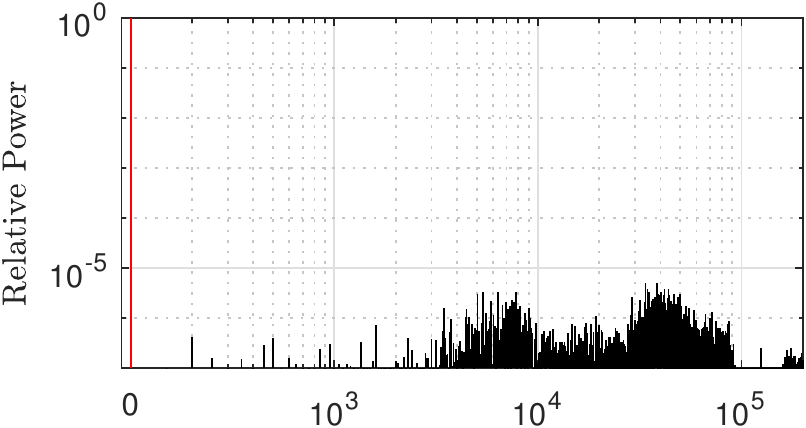}
	\caption{Distortion spectrum for noise allowed below \SI{10}{\kilo\hertz} and above \SI{28}{\kilo\hertz}, while the noise in between is reduced, demonstrating the flexibility of the control scheme to shape the output spectrum.}
	\label{fig:letsrock_spec}	
\end{figure}

\begin{figure}
	
	\begin{subfigure}{\columnwidth}
	\centering
	\includegraphics{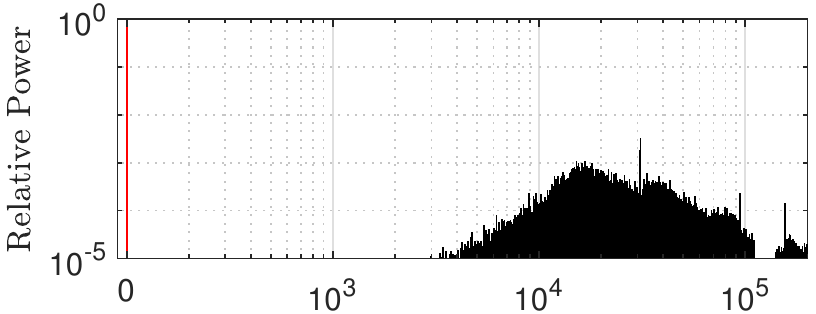}
		\subcaption{\SI{22.5}{\kilo\hertz} average switching frequency}
		\label{fig:sf_red1}
	\end{subfigure}\\
	\begin{subfigure}{\columnwidth}
		\centering
		\includegraphics{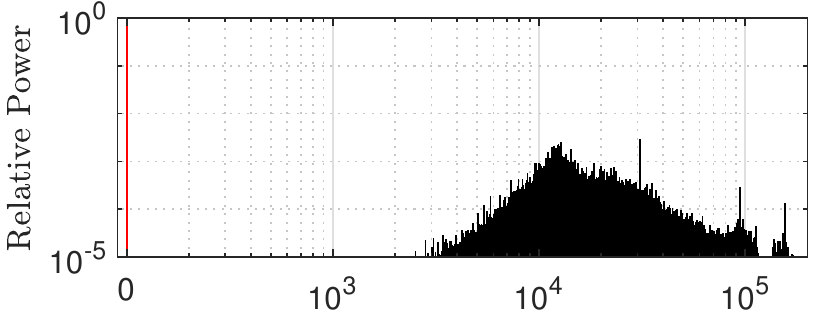}
		\subcaption{\SI{13.5}{\kilo\hertz} average switching frequency}
		\label{fig:sf_red2}	
	\end{subfigure}
	\begin{subfigure}{\columnwidth}
		\centering
		\includegraphics{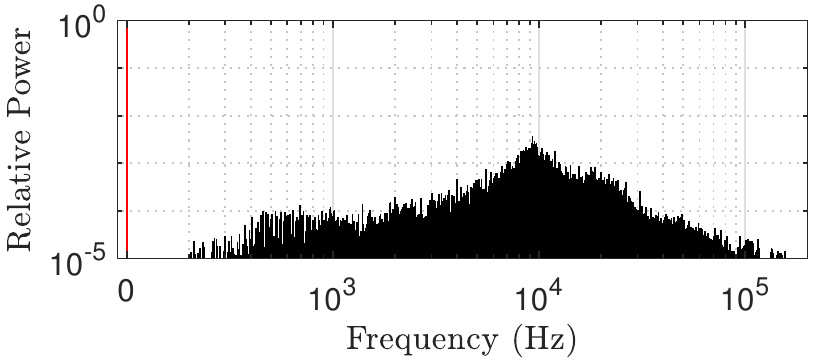}
		\subcaption{\SI{9.1}{\kilo\hertz} average switching frequency}
		\label{fig:sf_red3}
	\end{subfigure}
	\caption{Output spectrum of the prototype as the average switching frequency is decreased using the weight $\lambda_2$. A significant reduction in average switching frequency from \SI{22.5}{\kilo\hertz} to \SI{13.5}{\kilo\hertz} is achievable without greatly impacting the shape of the distortion.}
	\label{fig:sf_red}
\end{figure}

\begin{figure}
	\centering
	\includegraphics{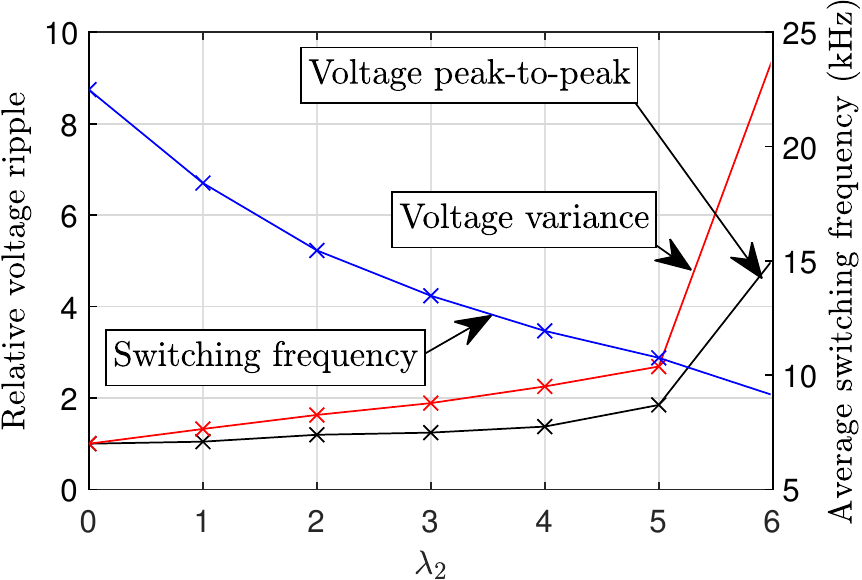}
	\caption{Trade-off between the voltage ripple of the converter and the average switching frequency. As $\lambda_2$ is increased, the average switching frequency reduces while the voltage ripple increases. At $\lambda_2=6$, the voltage ripple increases drastically.
		}
	\label{fig:swpen_stat}	
\end{figure}

\subsection{Results}

\begin{table}
	\centering
	\caption{This table shows the influence of the parameter $K_\mathrm{max}$ on the average switching frequency of the converter and the output voltage ripple. The voltage ripple can be greatly reduced at the cost of a slight increase in average switching frequency.}
	\label{tab:ripred}
	\begin{tabular}{c|c|cc}
		\multirow{2}{*}{$K_\mathrm{max}$} &Avg. switching		&\multicolumn{2}{c}{Relative voltage ripple}\\
					& 				rate		& Variance				& peak-to-peak \\ \hline
		$\infty$	& \SI{9.13}{\kilo\hertz}	& {3.66}	& {2.35}\\
		$20$       & \SI{9.55}{\kilo\hertz} 	& {1.35}	& {1.22}\\
		$10$ 		& \SI{13.0}{\kilo\hertz} 	& {1.00}	& {1.00}\\

	\end{tabular}
\end{table}

\begin{figure*}
\begin{center}
	\includegraphics{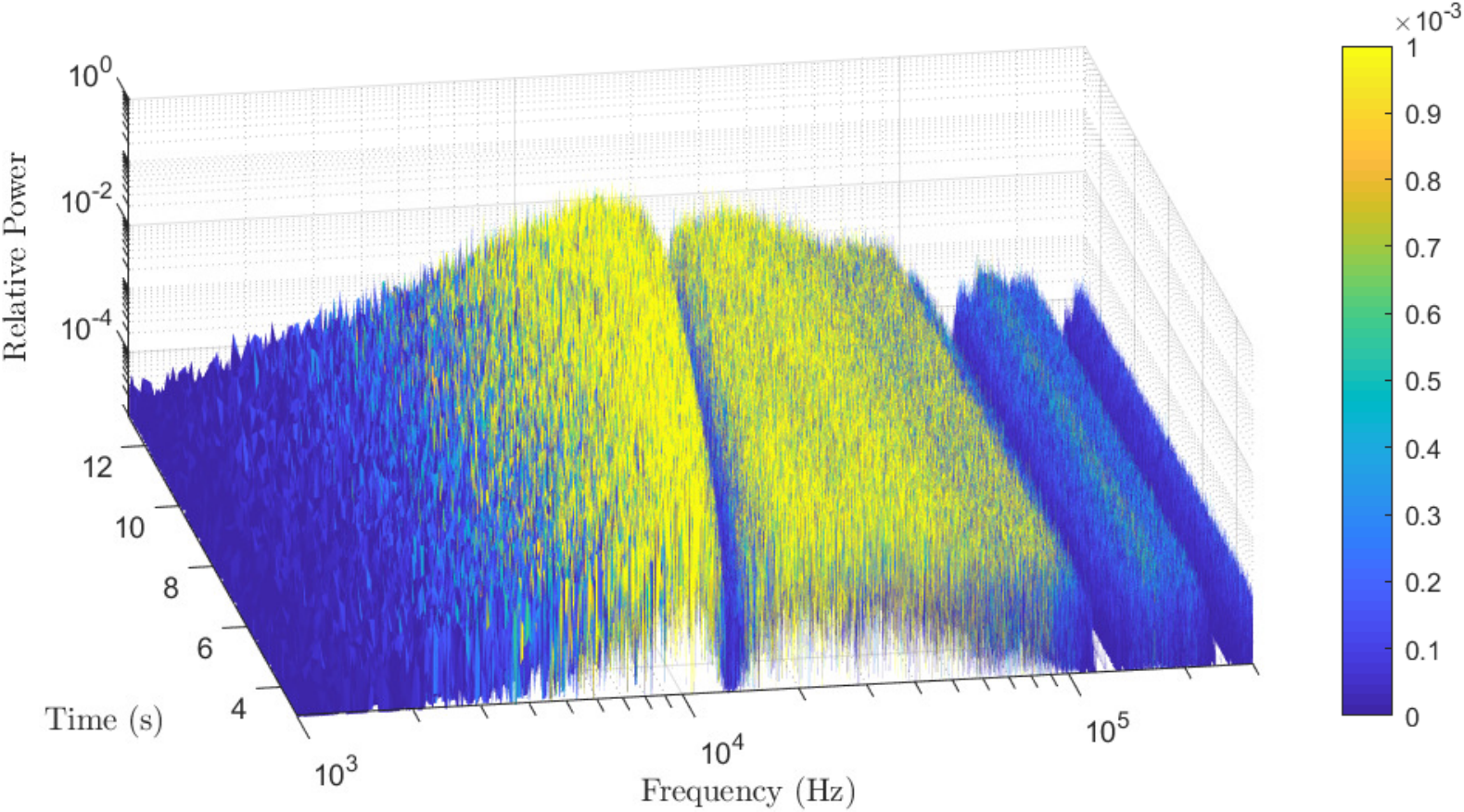}
	\end{center}
	\caption{Spectrogram of the output of the controller demonstrating the ability of the spectral control to change the output spectrum shape in real-time. A spectral gap is implemented at \SI{10}{\kilo\hertz} and shifted to \SI{23}{\kilo\hertz} during operation.}
	\label{fig:STFT}
\end{figure*}

\cref{fig:timemeas} illustrates the output current and voltage of the presented spectral converter under an output load step from \SI{5}{\ampere} to \SI{10}{\ampere}. The output voltage is maintained at the reference \SI{12}{\volt} while the current increases. {The maximum output peak-to-peak voltage ripple in this operating point is \SI{4.4}{\ampere}.}
{While the transient performance of the output controller is not the focus of the paper, the use of different control methods such as optimal control or state-space control could further improve the dynamics of the converter.}

\cref{fig:timespecmeasbef,fig:timespecmeasaft} show the output voltage spectrum of the converter before and after the load step. For this measurement, two gaps in the are generated in the distortion spectrum, at \SI{15}{\kilo\hertz} and \SI{20}{\kilo\hertz}. These gaps have a width of ca.\ \SI{2}{\kilo\hertz} and a depth of over \SI{30}{\decibel}, keeping these frequency ranges practically free of \gls{EMI}. Both before and after the load step, the output spectrum adheres closely to the reference spectrum, proving that the proposed method is well-suited for both stationary and dynamic load cases.

To demonstrate the flexibility of the spectral control scheme, \cref{fig:letsrock_spec} presents another example of an output distortion shape that can be generated. The spectral controller generates two regions of increased noise while permitting limited noise over a wide band from \SIrange{10}{28}{\kilo\hertz}. \cref{fig:timespecmeas,fig:letsrock_spec} are generated using the same settings apart from the spectral filter \G.


\cref{fig:sf_red} depicts the output spectrum of the converter as the average switching rate is reduced by increasing weight $\lambda_2$. The reference spectrum is a distortion spectrum with a constant distortion power from \SI{10}{\kilo\hertz} to {half the update frequency (\SI{62.5}{\kilo\hertz}).}
The average switching rate is reduced from \SI{22.5}{\kilo\hertz} in \cref{fig:sf_red1} where $\lambda_2=0$, to \SI{13.5}{\kilo\hertz} ($\lambda_2=3$) in \cref{fig:sf_red2} and finally \SI{9.1}{\kilo\hertz} in \cref{fig:sf_red3} ($\lambda_2=6$). 

{As a consequence of the decrease in average switching frequency, the output voltage ripple increases as $\lambda_2$ increases, as shown in \cref{fig:swpen_stat}. The voltage ripple is evaluated using the variance of the output voltage as well as the maximum peak-to-peak ripple. \cref{fig:swpen_stat} demonstrates that the average switching frequency can be reduced significantly from \SIrange{22.5}{10.8}{\kilo\hertz} while the output voltage ripple increases by a factor of 1.8 (peak-to-peak) and 2.7 (variance), exemplifying the trade-off between average switching frequency and output ripple. The output \gls{SFDR} of all measurements is similar, within \SIrange{22.8}{25.5}{\decibel}.}

This behavior highlights that the term $\lambda_2$ in the predictive cost function can be used to significantly reduce the switching frequency and therefore the switching losses of the \gls{DCDC} converter, whose switching losses are proportional to the switching frequency at a constant load.

An additional feature of the control scheme is the ability to control the voltage ripple of the converter, by enforcing a switching action if the spectral and switching rate optimization does not demand a switching action after a certain number of steps $K_{\mathrm{max}}$. {\cref{tab:ripred} demonstrates the effect of \Km\ on the voltage ripple and average switching frequency when $\lambda_2=6$. For this case, the variance of the output voltage ripple can be reduced by a factor of more than 2 while the average switching rate is increased by less than 5\%, from \SIrange{9.13}{9.55}{\kilo\hertz}.}

One of the advantages of the presented control scheme is the ability to adapt the control during operation to different conditions. This can be beneficial if the converter couples to changing acoustic resonances or sensitive electronics change their sensitive frequency band, e.g., a tuner, a bus with variable bus data rate, or thermally drifting resonances.

The ability of the control scheme to adapt during operation is exemplified in \cref{fig:STFT}, which shows a spectrogram of the output voltage as the spectral filter is changed. This could be used to keep a frequency free of noise during a radio tuner scan. Specifically, a spectral gap is implemented at \SI{10}{\kilo\hertz} and then shifted to \SI{23}{\kilo\hertz} at a rate of \SI{1.2}{\kilo\hertz\per\second}. The depth of the spectral gap throughout the spectrogram is circa a factor of 100 compared to the encompassing bands.

{The experimental verification of the spectral control method shows that the model predictive spectral control can shape the output spectrum of a \gls{DCDC} converter at run time while actively controlling the average converter switching rate and the output voltage ripple. The benefit of this control scheme compared to conventional spectrum spreading methods---such as frequency modulation---is that the predictive nature allows the user to intuitively design the output spectrum using only the filter \G, without the need to consider how a large number of variables affect the spectrum, such as the speed of the frequency modulation, the range of the frequency modulation, jumps in the modulation frequency the type of frequency modulation. Furthermore, the average switching rate and output voltage ripple can be combined in the cost function to a single control law.

Disadvantages of the spectral control method are the fact that significant computational resources are necessary to calculate the \gls{DFT} for all output options in real time, and that the average frequency and output voltage ripple of the converter cannot be determined in advance. These need to be tuned heuristically using the weights in the cost function.
}

\section{Conclusion}

This paper presented a model-predictive control method that allows the spectral distortion of a power-electronic converter to be controlled according to predefined criteria, while limiting the voltage ripple and regulating the switching rate. The control scheme was implemented on an \gls{FPGA} and verified using a \SI{48}{\volt} to \SI{12}{\volt} converter.

We demonstrated that with several artifices the control does not need to explicitly evaluate a single numerical Fourier transform so that the predictive spectral control operates at real time. Still, computational resources limit the spectral bandwidth that can be controlled and the achievable output voltage and current ripple levels. However, this proof of concept demonstrates that the output spectrum of an automotive \gls{DCDC} converter can be shaped according to a predefined shape allowing the emitted \gls{EMI} to be greatly reduced, and increasing size and speed of \gls{FPGA}s and \gls{DSP}s will increase the controllable bandwidth without further ado.

The control scheme is able to reduce the spectral density of the distortion power by a factor of 128 compared to conventional \gls{PWM} and, for instance, incorporate gaps in the spectral distortion to suppress interference at specific frequency bands. Furthermore we demonstrated that the shape of the output distortion can be changed on-the-fly, depending on the current requirements of the application.


\bibliographystyle{IEEEtran}
\bibliography{Transactions_Bib}


\clearpage

\end{document}